\documentclass[prl,twocolumn,superscriptaddress]{revtex4-1}

\usepackage{color}
\usepackage[english]{babel}
\usepackage[T1]{fontenc}
\usepackage[latin9]{inputenc}
\usepackage{latexsym}
\usepackage{float}
\usepackage{amsmath}
\usepackage{graphicx}
\usepackage{times}   
\usepackage{esint}
\usepackage[unicode=true,pdfusetitle,
 bookmarks=false,colorlinks=true,citecolor=blue,urlcolor=blue,linkcolor=red]{hyperref}

\makeatletter

\@ifundefined{textcolor}{}
{%
 \definecolor{BLACK}{gray}{0}
 \definecolor{WHITE}{gray}{1}
 \definecolor{RED}{rgb}{1,0,0}
 \definecolor{GREEN}{rgb}{0,1,0}
 \definecolor{BLUE}{rgb}{0,0,1}
 \definecolor{CYAN}{cmyk}{1,0,0,0}
 \definecolor{MAGENTA}{cmyk}{0,1,0,0}
 \definecolor{YELLOW}{cmyk}{0,0,1,0}
}

\@ifundefined{date}{}{\date{}}
\AtBeginDocument{
  
}
\makeatother

\newcommand{\SAVE}[1]{}

\newcommand{\prlsec}[1]{\emph{#1---}}

\newcommand{\Jbq}{{J_{bq}}}
\begin{document}
\renewcommand\abstractname{}

\title{Trimerized ground state of the spin-1 Heisenberg antiferromagnet on the kagome lattice}
\author{Hitesh J. Changlani}
\affiliation{Department of Physics, University of Illinois at Urbana-Champaign, Urbana, IL 61801, USA}
\affiliation{Institute for Quantum Optics and Quantum Information of the Austrian Academy of Sciences, A-6020 Innsbruck, Austria}
\affiliation{Institut f\"ur Theoretische Physik, Universit\"at Innsbruck, A-6020 Innsbruck, Austria}
\author{Andreas M. L\"auchli}
\affiliation{Institut f\"ur Theoretische Physik, Universit\"at Innsbruck, A-6020 Innsbruck, Austria}

\date{\today}

\begin{abstract}
We study the phase diagram of the spin-1 quantum bilinear-biquadratic 
antiferromagnet on the kagome lattice, using exact diagonalization (ED) 
and the density matrix renormalization group (DMRG) algorithm. 
The $\mathrm{SU(3)}$-symmetric point of this model Hamiltonian is 
a spontaneously trimerized state whose qualitative nature persists 
even at the Heisenberg point, a finding that contrasts previous proposals. 
We report the ground state energy per site of the Heisenberg 
model to be $-1.410(2)$ and establish the presence of a spin gap. 
\end{abstract}

\maketitle
\prlsec{Introduction} The discovery of experimental realizations of kagome antiferromagnets~\cite{Nocera_Kagome_2007,YS_Lee_2012} 
and indications that they have exotic ground states has spurred immense activity in the last few years. 
The nature of the ground state is unresolved for even 
the simplest realistic model, the nearest neighbor 
spin-1/2 kagome Heisenberg antiferromagnet (KHAF)~\cite{Zeng_Elser, 
Marston_Zeng, Lecheminant, Nikolic_kagome,Singh_Huse, Wen_kagome, MERA_kagome, White_kagome}. However, recent advances in numerical algorithms have enhanced our understanding of these systems~\cite{White_kagome, Depenbrock_Kagome, Jiang_Balents_Nature, Iqbal_U1, Clark_Kagome, Bauer_Kagome} 

In contrast to the spin $S=1/2$ case, little has been definitively established for the 
ground state of the $ S > 1/2$ case. When $S$ is large, 
as is the case for the $S = 5/2$ compound KFe$_3$(OH)$_6$(SO$_4$)$_2$~\cite{Matan}, 
long-range magnetic order of the $\sqrt{3}\times\sqrt{3}$ type 
is expected~\cite{Huse_Rutenberg,Henley_kagome}. However, for the intermediate 
spin case, $S=1$~\cite{Hida,Xu_Moore,Gotze} and $S=3/2$~\cite{Lauchli_kagome_3b2}, 
the theoretical situation is unclear. There exist several experimental 
motivations~\cite{Pati_Rao} for studying this problem. For example, 
KV$_3$Ge$_2$O$_9$~\cite{Narumi_Expt} and BaNi$_3$(OH)$_2$(VO$_4$)$_2$~\cite{Freedman_Expt} 
are candidates for $S=1$, and the chromium-jarosite (KCr$_3$(OH)$_6$(SO$_4$)$_2$) 
has been reported to be a $S=3/2$ kagome antiferromagnet~\cite{Cr_Jarosite_Expt}. 

The focus of this Rapid Communication is the $S=1$ case, with emphasis on the KHAF. 
Previous numerical studies of the $S=1$ XXZ model with on-site 
anisotropy~\cite{Damle_Senthil,Isakov_Kim_2009} have 
shed light on the phase diagram, but the approach is limited for the KHAF. 
Recent coupled cluster calculations~\cite{Gotze} show 
that the $S=1$ KHAF has no long-range magnetic order, 
in contrast to previous analytic results~\cite{Xu_Moore}. Thus, 
the definitive characterization of the ground state 
remains an open question.

\begin{figure*}[htpb]
\centering
\includegraphics[width=1\linewidth]{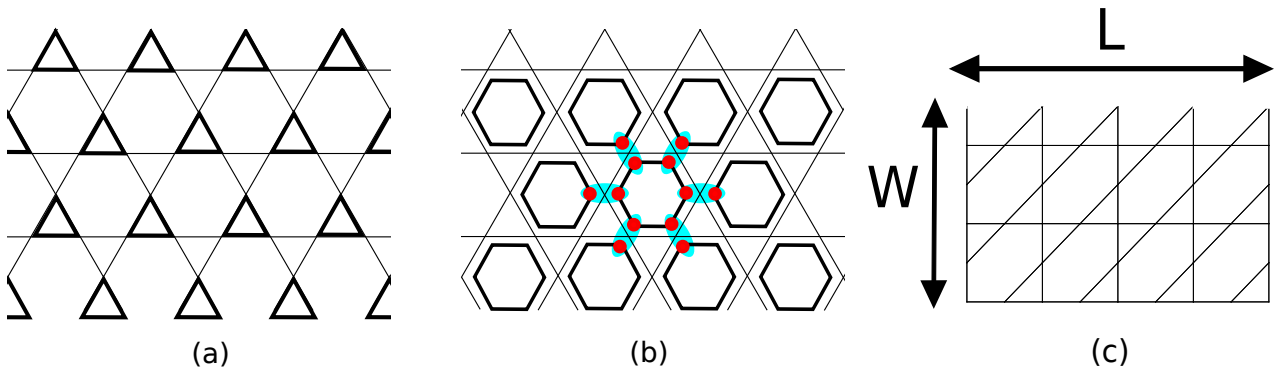}
\caption{(Color online) (a) shows a schematic of the simplex solid on the kagome lattice. 
The bond thicknesses represent the relative magnitude of the bond energy. 
(b) shows a schematic of the Hexagon Singlet Solid (HSS). 
Each spin-1 (depicted in blue) fractionalizes into two spin-1/2 (shown by red circles). 
The spin-1/2's on the hexagons form a singlet, shown by the black lines connecting them.
(c) shows the cylindrical geometry used in the DMRG calculation. 
Periodic boundary conditions in the width direction have not been shown.
}
\label{fig:kagome} 
\end{figure*}

Based on exact diagonalization (ED) of the $S=1$ KHAF, Hida proposed that 
the ground state is a Hexagonal Singlet Solid (HSS) with a spin gap~\cite{Hida}. 
The HSS is a translationally invariant state that is described by an Affleck-Kennedy-Lieb-Tasaki 
(AKLT)~\cite{AKLT} type wavefunction. As is schematically 
depicted in Fig.~\ref{fig:kagome}(b), all the spin-1's fractionalize into 
two spin-1/2's and then the spin-1/2's on every hexagon form a singlet state. 
However, a recent experiment~\cite{Awaga} with m-MPYNN-BF$_4$, 
believed to be a $S=1$ KHAF, has observed magnetization plateaus different from those predicted by 
the HSS phase~\cite{Hida_Plateau}, calling for a review of this picture. 

In this Rapid Communication, we use ED and the 
density matrix renormalization group (DMRG) algorithm~\cite{dmrg_white} 
for cylindrical geometries~\cite{2D_DMRG}. We show that even though 
the HSS has a competitive energy ($\approx -1.36$ per site) 
in comparison to the DMRG results ($\approx -1.41$ per site), the qualitative picture 
obtained from the latter is that of a trimerized ground state, schematically 
illustrated in Fig.~\ref{fig:kagome}(a). This state, 
referred to as the simplex-solid~\cite{Arovas_Simplex} 
or simplex-valence bond crystal, is a symmetry broken state 
where the three spin-1's living on each up (or equivalently down) 
pointing triangles form collective singlets or "trimers". 

We find no long-range spin-spin 
correlations and a finite spin gap of $\sim 0.2 - 0.3 $, 
for the choice of lattice geometries studied. In addition, 
the energy of a recently proposed ground state 
candidate $Z_2$ spin liquid, the Resonating AKLT state (RAL)~\cite{Kagome_spin_1_PEPS}, 
is found to be higher than both the HSS and the trimerized state found in DMRG. 
 
We have considered the phase diagram of 
the nearest neighbor bilinear-biquadratic model,
\begin{equation}
\mathcal{H}=J_{bl} \sum_{\langle ij\rangle}{\mathbf{S}_{i}\cdot{\mathbf{S}_{j}}} + J_{bq}\sum_{\langle ij\rangle} \left({\mathbf{S}_{i}} \cdot{\mathbf{S}_{j}} \right)^{2} 
\label{eq:hamiltonian}
\end{equation}
where $\langle ij\rangle$ refer to nearest neighbor pairs, 
$J_{bl}$ is the bilinear Heisenberg coupling 
(set to $J_{bl}=1$), and $\Jbq$ is the biquadratic coupling. 
While a previous tensor network study showed 
the ground state to be a simplex solid at the $\mathrm{SU(3)}$ 
symmetric point ($J_{bl}=\Jbq$)~\cite{Corboz_Kagome_simplex}, 
here we provide evidence that this trimerization survives on 
reducing the magnitude of $\Jbq$ all 
the way to zero. A quantum phase transition to a 
ferroquadruolar spin nematic is observed only at 
$\Jbq \sim -0.16$. 

\prlsec{The Heisenberg point}
We consider $\Jbq=0$, the Heisenberg point, and assess 
the quality of the HSS wavefunction with respect to ED 
calculations. Following Hida~\cite{Hida}, we associate two spin-1/2 degrees of freedom 
with every spin-1, and define,
\begin{equation}
 |+1 \rangle \equiv \frac{\psi_{1/2,1/2}}{\sqrt{2}}\;\;\;\; |0\rangle \equiv \psi_{1/2,-1/2} \;\;\; |-1 \rangle \equiv \frac{\psi_{-1/2,-1/2}}{\sqrt{2}}
\end{equation}
where $\psi_{\alpha,\beta}\equiv \frac{1}{\sqrt{2}} \left ( \psi_{\alpha}\psi_{\beta} + \psi_{\beta}\psi_{\alpha} \right )$ 
and where $\alpha$,$\beta$ correspond to spin-1/2 degrees of freedom and have value $\pm 1/2$. 
Then the HSS wavefunction is defined to be, 
\begin{equation}
	\phi_{\text{HSS}} = \otimes \psi_{\alpha_i,\beta_i} \; \prod_{i} (\delta_{\alpha,\gamma_i}+ \delta_{\beta,\gamma_i}) \prod_{p} w^{\gamma_{i_p},\gamma_{j_p},k_p,l_p,m_p,n_p}
\end{equation}
where $p$ is a label used to distinguish the hexagons and $i_p$ through $n_p$ refer to the sites on the elementary hexagon 
(in contiguous order) and $\gamma_{i_p}$ through $\gamma_{n_p}$ correspond to the spin-1/2 degrees on those sites. 
$w^{\gamma_{i_p},\gamma_{j_p},\gamma_{k_p},\gamma_{l_p},\gamma_{m_p},\gamma_{n_p}}$ is the coefficient of the 
lowest energy singlet state of a $S=1/2$ Heisenberg model on a hexagon.

Table I shows the energy of the HSS, the RAL~\cite{Kagome_spin_1_PEPS} 
and ground state wavefunctions from ED for various finite clusters 
with periodic boundary conditions (the geometries and nomenclature are 
the same as Ref.~\cite{Hida}). We estimate the HSS energy in the thermodynamic limit 
to be $\approx -1.36$ per site~\footnote{The 
energy of the HSS wavefunction is found to be lower than that 
reported previously~\cite{Hida}}. 
This is comparable to the exact energy of $\approx -1.4$, 
and much lower than the RAL energy, suggesting that 
the HSS is a competitive candidate for the ground state. 

However, a clear picture of the ground state emerges only for 
larger systems, which were studied with DMRG. 
Cylinders with periodic boundaries 
in the short (width) direction and open boundaries 
in the long (length) direction, as shown in Fig.~\ref{fig:kagome}(c), 
were chosen for the simulations and finite size analyses. 
In order to have complete hexagons, even widths were chosen.

\begin{table}[htpb]
\begin{center}
\begin{tabular}{|c|c|c|c|c|c|}
\hline
Wavefunction            & 12            & 15             &  18 $a$     & 18 $b$            & $\infty$              \tabularnewline
\hline
\hline
HSS                     & -1.38781      & -1.36024       & -1.36108    & -1.36995          & $\approx$ -1.36       \tabularnewline
\hline
RAL~\cite{Kagome_spin_1_PEPS}   & -     & -              & -           & -1.38             & -1.2696               \tabularnewline
\hline
ED                      & -1.46841      & -1.44958       & -1.45110    & -1.43926          & $\approx$ -1.4        \tabularnewline
\hline
\end{tabular}
\caption{Energy per site for the Hexagon Singlet State (HSS), Resonating AKLT state (RAL) and 
exact diagonalization (ED) wavefunctions on various kagome clusters with periodic boundary conditions.}
\end{center}
\end{table}

\begin{figure*}[htpb]
\centering
\includegraphics[width=1\linewidth]{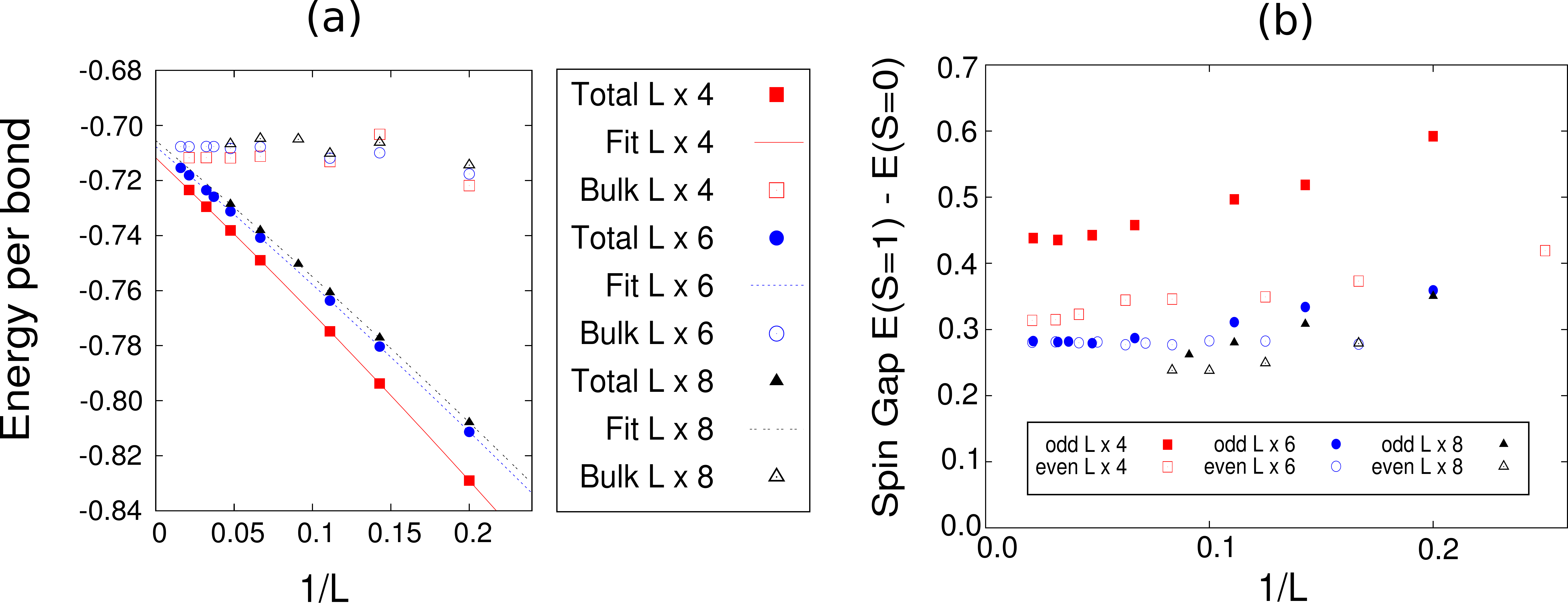}
\caption{(Color online): (a) The total ground state energy per bond 
for cylinders of odd lengths and different widths 
is extrapolated to infinite length by fitting to 
the functional form, Eq.~\eqref{eq:e_vs_L}. The bulk energy 
(see text) is also shown. (b) shows the spin gap for various cylinder 
widths and lengths. The estimated gap in the 
infinite length limit is finite.}
\label{fig:figure_2} 
\end{figure*}	

The number of renormalized states (denoted by $m$) kept 
in the DMRG simulations, were typically $2000$, $3000$ and $4000$ 
for widths $4$, $6$ and $8$ respectively. On cylinders with 
widths $4$ and $6$, and odd lengths (these have equal numbers 
of up and down pointing triangles), a pattern of alternating strong 
and weak trimers propagates from both the left and right edges. 
These competing patterns superpose in the center of the finite sample, 
leading to uniform bond energies (the bond energy being defined as 
$\langle \mathbf{S}_i \cdot \mathbf{S}_j \rangle$ for $i,j$ 
being nearest neighbor sites). On the even-length cylinders 
(with more down triangles than up), the left-most row of boundary 
sites form dimers, effectively decoupling them from the bulk of the system. 
Thus, the even-length cylinders have similar bulk properties 
to the odd-length cylinders. 

For width $8$ cylinders, the tendency to form dimers along the width direction 
is suppressed and a robust trimerization pattern is observed throughout the bulk. 
For the odd lengths, DMRG tends to break the symmetry between the up 
and down pointing triangles, which we take to be evidence that the system 
prefers to trimerize. This is a "finite $m$" effect, as an 
\emph{exact} calculation should yield a perfect superposition of both trimer states.

To estimate the energy per bond 
in the thermodynamic limit, we used two different procedures. 
First, we consider the energy $E$ of 
the entire sample comprising of $N_{b}$ bonds, 
and fit it to the functional form,
\begin{equation}
	E(L,W)/N_{b}(L,W) = E_0 + a_1/L + a_2/L^{2} \label{eq:e_vs_L}
\end{equation}
where $N_b(L,W)$ is the number of bonds and $e_b$,$a_1$,$a_2$ are fit parameters. In the second method, we average the bond energies on a central feature, 
such as the bowtie or "star" consisting of three up and three down triangles. 
We refer to this estimate as the "bulk" energy. 
Figure~\ref{fig:figure_2}(a) shows the 
length dependence of the energy and its extrapolation 
to infinite length for different cylinder widths. Both 
analyses yield similar estimates; for the width $4$, $6$ and $8$ 
cylinders the values of the energy per bond are 
$-0.7117(1)$, $-0.7067(1)$ and $-0.7058(4)$ respectively. 
Assuming small variations for energy estimates beyond $W > 8$, 
the energy per bond in the thermodynamic limit is $-0.705(1)$, 
which in terms of the energy per site is $-1.410(2)$. This is 
comparable to (and slightly lower than) extrapolated coupled 
cluster results~\cite{Gotze} ($E_0=-1.4031$)~\footnote{The 
reported series expansion estimate of $E_0=-1.4468$~\cite{Hida} does not agree with our DMRG result.}.

Next, we verified the presence of a spin gap in the thermodynamic limit, 
by calculating the energy difference between the singlet and triplet 
states for both even and odd length 
cylinders. Our results are shown in Fig.~\ref{fig:figure_2}(b). 
The magnetization of the first excited state is distributed 
over the entire sample, establishing that the excitation is a bulk one. 
The large variation in the energy gap for the width $4$ and 
the other larger cylinders is a finite size effect; this qualitative 
difference is also seen in ground state energy estimates. 
The trends for width $6$ and $8$ indicate that the spin gap is in the range $0.2 - 0.3 $. 

To build further confidence in these results, 
we study the bilinear-biquadratic (BLBQ) model~\eqref{eq:hamiltonian} 
and use $\Jbq$ as a knob to connect the Heisenberg point 
to the $\mathrm{SU(3)}$ point. 
Analyzing other Hamiltonians should lead to similar conclusions. 
For example, an extended-range Heisenberg model, studied by 
Cai et al.~\cite{CCW}, also has a trimerized ground state.

\begin{figure}[htpb]
\centering
\includegraphics[width=1\linewidth]{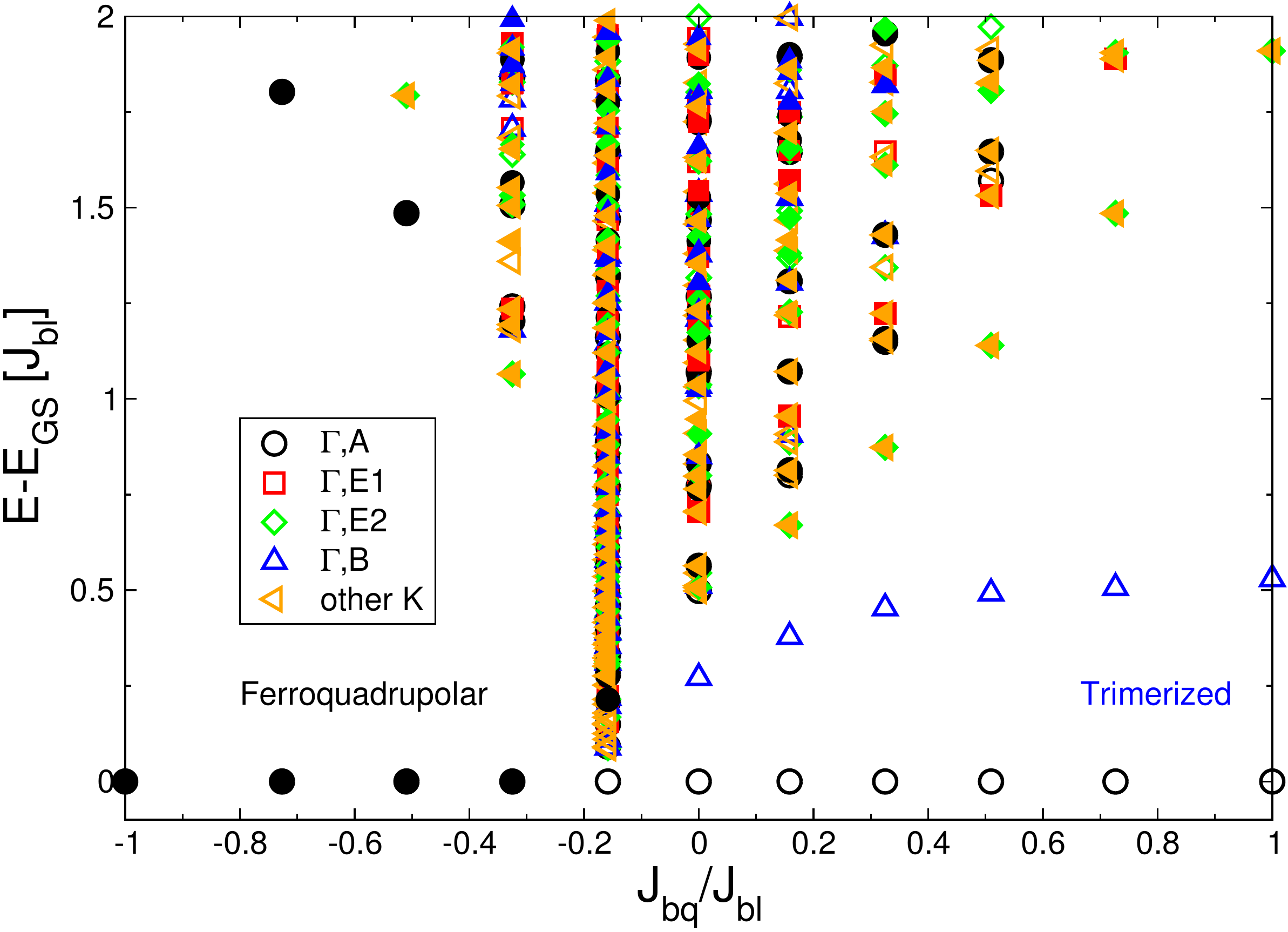}
\caption{(Color online): Low-energy 
spectrum of the BLBQ model on the $21$ site kagome lattice, 
resolved by lattice momenta, as a function of $\Jbq$. 
On tuning $\Jbq$ from $1$ towards $0$, the 
low-energy features appear adiabatically connected, suggesting 
the persistence of the trimerized phase to the 
Heisenberg point. Qualitative changes in the 
energy spectrum seen at a negative value of $\Jbq$, 
indicate a quantum phase transition to a ferroquadrupolar phase.}
\label{fig:ed_blbq_energies} 
\end{figure}

\prlsec{The bilinear-biquadratic (BLBQ) model}
\label{sec:blbq_model}
For insights into the BLBQ model, we 
performed ED calculations on a $21$ site sample with 
periodic boundary conditions. Multiple low-energy 
excited state energies, resolved by spatial momenta, 
have been plotted in Fig.~\ref{fig:ed_blbq_energies}. 
On tuning $\Jbq$ from $1$ towards $0$, 
we find no energy crossings in the first few 
states in the low-energy manifold. In the range 
$ - 0.2 < \Jbq < - 0.1 $, a marked decrease in energy spacings 
(or increased crowding of energy levels) 
and the appearance of a small finite size gap, are indicative 
of a quantum phase transition. 

\begin{figure}[htpb]
\centering
\includegraphics[width=1\linewidth]{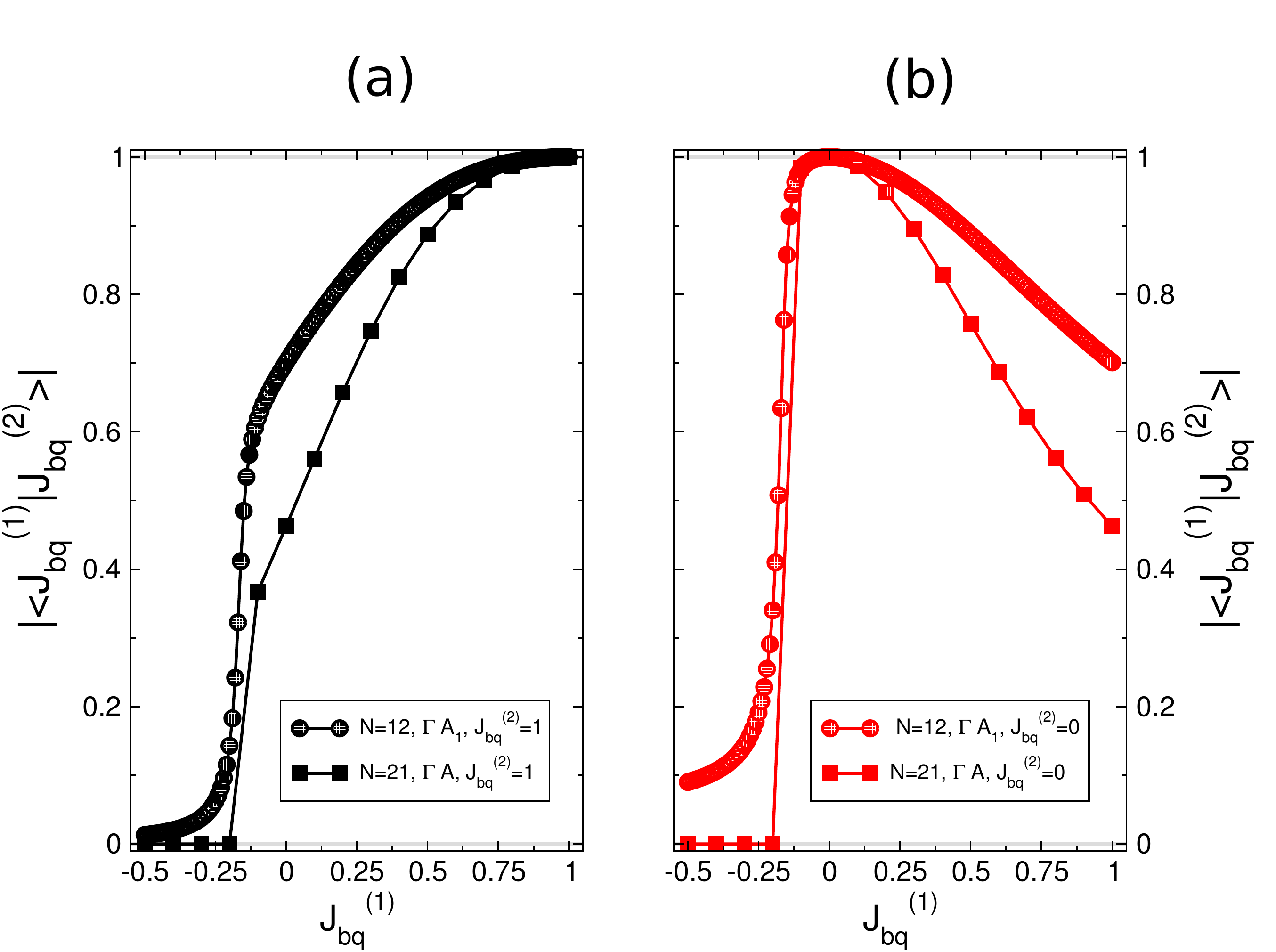}
\caption{(Color online): Fidelity (overlap) of ground state wavefunctions for various values of 
$\Jbq$ with respect to a reference wavefunction for the $12$ and $21$ site kagome clusters. 
The reference wavefunction is chosen to be the ground state of 
(a) the $\mathrm{SU(3)}$ symmetric model 
known to favor a trimerized (simplex solid) phase 
and (b) the Heisenberg model whose qualitative nature remains to be established 
and is the subject of this study. The abrupt change in fidelity 
is found to occur in both cases around $ -0.2 <\Jbq < -0.13$. 
}
\label{fig:fidelity} 
\end{figure}	

\begin{figure}[htpb]
\centering
\includegraphics[width=1\linewidth]{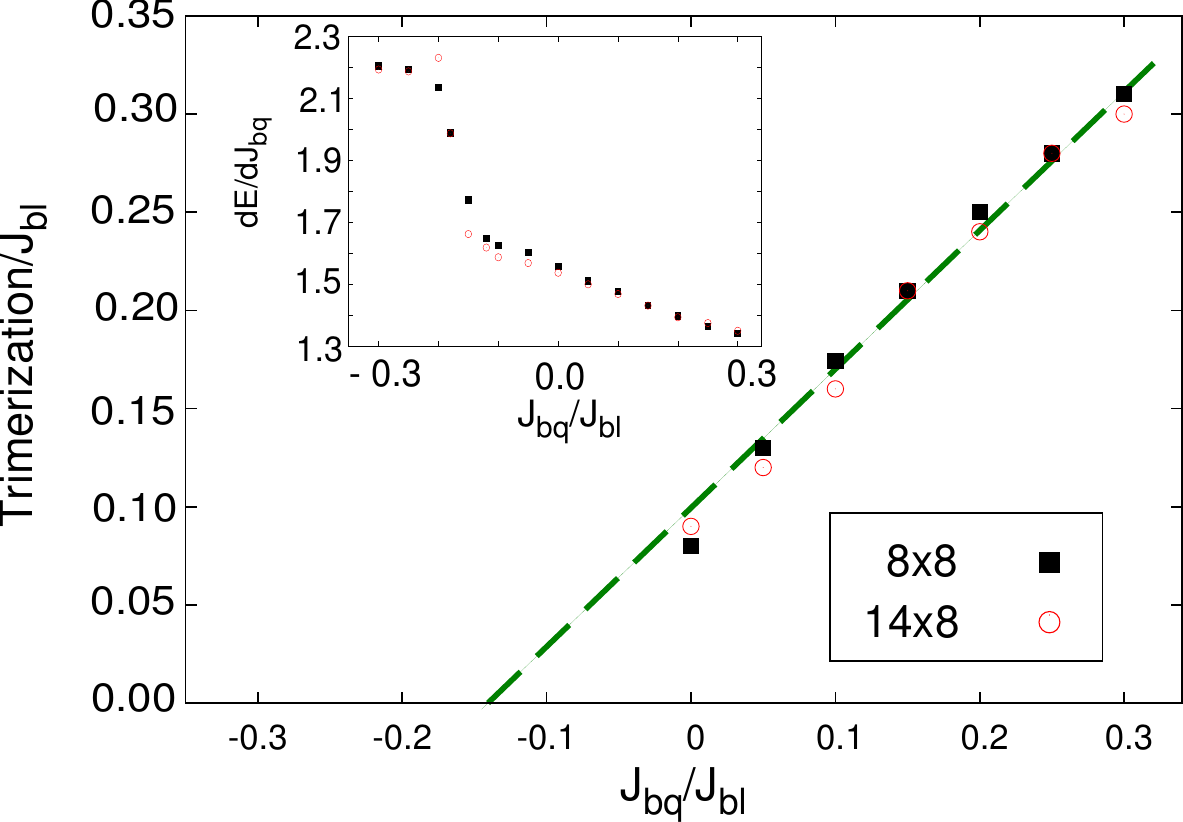}
\caption{(Color online): The trimerization order parameter (obtained 
by taking the difference of bond energies of neighboring triangles) 
as a function of $\Jbq$ for the $8 \times 8$ and 
$14 \times 8$ kagome lattice. The dashed line gives 
an extrapolated estimate of the $\Jbq^{*}$ at which the trimerization 
vanishes. Inset: Derivative of the total energy (per bond) 
with respect to $\Jbq$ shows an abrupt change around 
the same value of $\Jbq^{*}\approx-0.16$.} 
\label{fig:trimer} 
\end{figure}	

Next, we look for signatures of possible phase 
transitions as a function of $\Jbq$ by monitoring 
the wavefunction fidelity~\cite{Zanardi_fidelity}, defined as 
$F \equiv \langle \psi (p) | \psi_{ref} \rangle$, where 
$|\psi (p) \rangle$ is a wavefunction dependent on parameters $p$ 
and $|\psi_{ref}\rangle$ is a reference wavefunction. 
Fig.~\ref{fig:fidelity} shows fidelities of the $12$ and $21$ site clusters 
(as a function of $\Jbq/J_{bl}$) by fixing the reference wavefunction to 
be the ground state wavefunction of (a) 
the $\mathrm{SU(3)}$ model (Fig.~\ref{fig:fidelity}(a)) 
and (b) the Heisenberg model (Fig.~\ref{fig:fidelity}(b)). 
In either case, the fidelity decreases on going away from 
the chosen reference point and with increasing 
lattice size; the latter is expected because overlaps 
involve the multiplication of an increasing number 
of factors less than $1$. We consider an overlap of $0.45$ (for the $21$ site lattice) 
between the Heisenberg and $\mathrm{SU(3)}$ symmetric point wavefunctions
to be large and view the sharp fall 
in fidelity in the range $-0.2<J_{bq}< -0.13$ to be the only 
sign of a phase transition; we thus infer that 
the Heisenberg point corresponds to a trimerized ground state. 

The inferences from ED are verified on larger samples 
using DMRG, by considering a variety of metrics. 
First, as the inset of Fig.~\ref{fig:trimer} shows, 
the energy as a function of $\Jbq$ shows a discontinuity in its derivative 
at a value $\Jbq\approx-0.16$. This value coincides with the location of 
the minimum of the singlet-singlet gap, obtained by taking the 
energy difference of the lowest $S_z=0$ states in the DMRG method (not shown in plot). 
However, the most direct evidence is that of a non-zero trimer order parameter, 
defined to be 
\begin{equation}
	\text{Trimerization} \equiv \Big| \langle {\mathbf{S}_{i}\cdot{\mathbf{S}_{j}}} \rangle_{\Delta}- \langle {\mathbf{S}_{i}\cdot{\mathbf{S}_{j}}} \rangle_{\nabla} \Big|
\end{equation}
where $\langle {\mathbf{S}_{i}\cdot{\mathbf{S}_{j}}} \rangle_{\Delta (\nabla)}$ is the average spin-spin 
bond correlator on the up (or down pointing triangle). 
The trimerization is (relatively) uniform throughout the sample 
on the width $8$ cylinders: this data is used to 
determine the critical $\Jbq^{*}$ at which the phase transition occurs. 
When $\Jbq$ is close to $\Jbq^{*}$, the trimerization is small and inhomogenous 
and the presence of the open boundaries becomes important. 
This is why we used only the values of trimerization 
for $\Jbq \geq 0$ and extrapolated them to $\Jbq<0$ in Fig.~\ref{fig:trimer}. 

Below $\Jbq/J_{bl} \lesssim -0.16$, a ferroquadrupolar spin nematic is seen, 
a generic occurence in many $S=1$ antiferromagnets with negative 
biquadratic couplings~\cite{Penc_Lauchli}. This state 
has $\langle {\bf S_{i}}\rangle = 0$ but still breaks 
the spin rotational symmetry. This is verified by the observation 
that $\langle S_i^{+} S_i^{-} \rangle \neq \langle (S_i^{z})^{2} \rangle $ and 
that $ \langle (S_i^{z})^{2} \rangle$ abruptly changes from $0.66 (= 2/3)$ to 
$ \approx 0.4$ at the critical point. 

\prlsec{Conclusion} 
We have performed ED and DMRG calculations on 
the spin-1 kagome antiferromagnet with bilinear and biquadratic terms. 
We find evidence for trimerization at the Heisenberg point, which is not consistent with 
the hexagonal-singlet state (HSS) picture~\cite{Hida}, nor with the $\sqrt{3}\times\sqrt{3}$ 
order predicted by $1/S$ methods~\cite{Xu_Moore}. We also estimated 
the location of the phase transition from the trimerized state 
to the spin-nematic phase to be $\Jbq^{*} \sim -0.16$. 

Recently, Li et al.~\cite{Kagome_spin_1_PEPS} 
proposed a spin liquid ground state for the spin-1 KHAF, the 
resonating AKLT state (RAL), obtained by creating 
a uniform superposition of all possible "AKLT-loops". 
On an $18$ site lattice, the RAL energy is marginally 
lower than that of the HSS but in the infinite lattice limit 
is significantly higher~\cite{Kagome_spin_1_PEPS}. 
A plausible reason is that the RAL is dominated by 
long loops, that are still relatively short on an $18$ site lattice. 
Presumably, if the longest loops are penalized (i.e. a loop tension 
is added in the wavefunction), the RAL energy could improve significantly. 
Whether such a modification preserves the spin liquid properties or alternately 
drives it to a confining phase (such as the trimerized phase) is not known. 
Since the trimerization strength is small, it will thus be 
interesting to see if additional interactions at the Heisenberg point 
stabilize the RAL or HSS (or other exotic) states. 

Finally, we comment on the possible experimental consequences of our finding. 
Since trimerization does not change the magnetic unit cell structure of the kagome lattice, 
we still expect to see the 1/3 magnetization 
plateau for the $S=1$ KHAF (based on the Oshikawa-Yamanaka-Affleck criterion~\cite{Oshikawa}). 
However, prominent magnetization plateaus seen in the experiment with m-MPYNN-BF$_4$ (which also has a 
slight $\sqrt{3}\times\sqrt{3}$ distortion)~\cite{Awaga} correspond to 1/2 and 3/4 which is indicative of a 
magnetic unit cell with 12 atoms. Thus, we intend to understand 
the effective low-energy Hamiltonian better to resolve this issue.  

\prlsec{Acknowledgement} HJC thanks Prof. Christopher Henley 
for his guidance and for collaboration on related work. 
We are grateful to Christopher Henley, 
Shivam Ghosh, Kedar Damle, Steven White, Tyrel McQueen, 
Michel Gingras, Bryan Clark, Victor Chua and Gil Young Cho for discussions. 
We also acknowledge useful correspondence with H.-H. Tu 
regarding Ref.~\cite{Kagome_spin_1_PEPS}. This work was supported 
by the Austrian Ministry of Science BMWF as part of the Konjunkturpaket II of the 
Research Platform Scientific Computing at the University of Innsbruck. 
HJC acknowledges support from SciDAC grant DOE FG02-12ER46875. 
Calculations were also done on the Taub campus cluster at UIUC/NCSA.

\bibliographystyle{apsrev4-1}
\bibliography{refs}

\end{document}